\documentclass[acmsmall,natbib%,anonymous%,review
]{acmart}

\usepackage{xspace}
\usepackage{amsmath}

\usepackage{multirow}
\usepackage{subcaption}
\usepackage{graphicx}	
\usepackage{booktabs}
\usepackage{xcolor}
\usepackage{paralist} % not in ACM TAPS allowed packages

\AtBeginDocument{%
  \providecommand\BibTeX{{%
    \normalfont B\kern-0.5em{\scshape i\kern-0.25em b}\kern-0.8em\TeX}}}

\setcopyright{rightsretained}
\copyrightyear{2026}
\acmYear{2026}
\acmDOI{XXXXXXX.XXXXXXX}

\acmJournal{TORS}
\acmVolume{?}
\acmNumber{?}
\acmArticle{?}
\acmMonth{12}

\newcommand{\modeldummyhighpop}{\texttt{HighPop-\allowbreak{}Heavy}\xspace}
\newcommand{\modeldummylowpop}{\texttt{LowPop-\allowbreak{}Heavy}\xspace}
\newcommand{\modeldummycali}{\texttt{Calibrated}\xspace}

\begin{document}

%%
%% The "title" command has an optional parameter,
%% allowing the author to define a "short title" to be used in page headers.
%\title[]{The Connection Between Liking and Popularity Calibration\\ in Music Recommendation}
\title[Popularity Calibration in Music Recommendation]{Robustness and User-Perceived Value of Popularity Calibration in Music Recommendation: A User Study}

%%
%% The "author" command and its associated commands are used to define
%% the authors and their affiliations.
%% Of note is the shared affiliation of the first two authors, and the
%% "authornote" and "authornotemark" commands
%% used to denote shared contribution to the research.
\author{Oleg Lesota}
\email{oleg.lesota@jku.at}
\orcid{0000-0002-8321-6565}
\affiliation{%
  \institution{Johannes Kepler University Linz%\\ (Institute of Computational Perception, Multimedia Mining and Search Group)
  }
  \city{Linz}
  \country{Austria}
}

\author{Gustavo Escobedo}
\email{gustavo.escobedo@jku.at}
\orcid{0000-0002-4360-6921}
\affiliation{%
\institution{Johannes Kepler University Linz%\\ (Institute of Computational Perception, Multimedia Mining and Search Group)
}
  \city{Linz}
  \country{Austria}
}

\author{Bruce Ferwerda}
\email{bruce.ferwerda@ju.se}
\orcid{0000-0003-4344-9986}
\affiliation{%
  \institution{J\"{o}nk\"{o}ping University}
  \department{Department of Computer Science and Informatics}
  \city{J\"{o}nk\"{o}ping}
  \country{Sweden}
}

\author{Simone Kopeinik}
\email{skopeinik@know-center.at}
\orcid{0000-0002-6440-7286}
\affiliation{%
  \institution{Know-Center GmbH}
  \city{Graz}
  \country{Austria}
}

\author{Dominik Kowald}
\email{dkowald@know-center.at}
\orcid{0000-0002-3913-2946}
\affiliation{
  \institution{Know Center Research GmbH and University of Graz}
  \city{Graz}
  \country{Austria}
}

\author{Elisabeth Lex}
\email{elisabeth.lex@tugraz.at}
\orcid{0000-0001-5293-2967}
\affiliation{%
  \institution{Graz University of Technology}
  \city{Graz}
  \country{Austria}
}

\author{Markus Schedl}
\email{markus.schedl@jku.at}
\orcid{0000-0003-1706-3406}
\affiliation{
  \institution{Johannes Kepler University Linz %(Institute of Computational Perception, Multimedia Mining and Search Group)
  and 
  Linz Institute of Technology
  %(AI Lab, Human-centered AI Group)
  }
  \city{Linz}
  \country{Austria}
}
\renewcommand{\shortauthors}{Anonymous}

%%
%% The abstract is a short summary of the work to be presented in the article.

\begin{abstract}
Popularity calibration in recommender systems has been studied both as a form of user-centered personalization and as an indicator of popularity bias. Most existing work evaluates calibration through offline metrics, often assuming that users prefer recommendation lists whose popularity distribution matches their historical consumption profile. However, user studies on calibration remain limited, and existing findings suggest that calibrated recommendations do not necessarily have a strong effect on user experience. Moreover, although prior work has shown that calibration metrics can correlate with users' perceptions of recommendation lists, the robustness of this relation remains unclear under different levels of item familiarity and incomplete user-history information.
In this work, we study the perceived value and measurement reliability of popularity calibration in music recommendation. We construct personalized track lists from users' recent listening histories and use a controlled naive recommender to create lists with different popularity compositions: \modeldummyhighpop, \modeldummylowpop, and \modeldummycali. We investigate whether users perceive differences between these lists, whether calibrated lists are preferred, how robust JSD-based popularity calibration is under different familiarity and history-availability conditions, and how computational popularity labels align with users' own popularity judgments. Our results show that users perceive differences in popularity composition, but do not clearly prefer calibrated lists. We further find that the relation between JSD and perceived popularity depends on item familiarity, list composition, and available user history, while computational and user-judged popularity labels only weakly align. These findings contribute to a more critical understanding of popularity calibration as both an offline metric and a user-facing construct.
\end{abstract}

%%
%% The code below is generated by the tool at http://dl.acm.org/ccs.cfm.
%% Please copy and paste the code instead of the example below.
%%
% \begin{CCSXML}
% <ccs2012>
%  <concept>
%   <concept_id>10010520.10010553.10010562</concept_id>
%   <concept_desc>Computer systems organization~Embedded systems</concept_desc>
%   <concept_significance>500</concept_significance>
%  </concept>
%  <concept>
%   <concept_id>10010520.10010575.10010755</concept_id>
%   <concept_desc>Computer systems organization~Redundancy</concept_desc>
%   <concept_significance>300</concept_significance>
%  </concept>
%  <concept>
%   <concept_id>10010520.10010553.10010554</concept_id>
%   <concept_desc>Computer systems organization~Robotics</concept_desc>
%   <concept_significance>100</concept_significance>
%  </concept>
%  <concept>
%   <concept_id>10003033.10003083.10003095</concept_id>
%   <concept_desc>Networks~Network reliability</concept_desc>
%   <concept_significance>100</concept_significance>
%  </concept>
% </ccs2012>
% \end{CCSXML}

% \ccsdesc[500]{Computer systems organization~Embedded systems}
% \ccsdesc[300]{Computer systems organization~Redundancy}
% \ccsdesc{Computer systems organization~Robotics}
% \ccsdesc[100]{Networks~Network reliability}

%%
%% Keywords. The author(s) should pick words that accurately describe
%% the work being presented. Separate the keywords with commas.

\maketitle

\section{Introduction}

Popularity calibration has been proposed as a user-centered way to study popularity bias in recommender systems~\cite{DBLP:conf/um/AbdollahpouriMB21,DBLP:journals/tors/SilvaJ26}. Instead of only measuring whether a system over-recommends popular items overall, calibration compares the popularity distribution of a user's past interactions with that of the recommendation list. This distinction is important because the same list can be calibrated for one user and miscalibrated for another: a list dominated by popular tracks may match the history of a mainstream-oriented listener, while it may be miscalibrated for a listener whose history contains more niche or long-tail music~\cite{DBLP:conf/um/AbdollahpouriMB21,kowald_study_2023,DBLP:conf/recsys/LesotaMRBKLS21}.

This framing makes popularity calibration attractive as a personalization and bias-mitigation concept, but it also introduces an important assumption: the user's observed listening history is treated as a meaningful target distribution for future recommendations. In practice, this target distribution depends on several design choices, including how item popularity is computed, how popularity bins are defined, and which parts of the user's history are available for calibration~\cite{DBLP:conf/um/AbdollahpouriMB21,kowald_study_2023,DBLP:conf/recsys/LesotaMRBKLS21}. These choices are not merely technical, because they affect what it means for a recommendation list to be considered calibrated and what kind of user preference the metric is assumed to represent.

A further challenge is item familiarity. Prior work on perceived popularity bias in music recommendation suggests that users may not always recognize computational differences in popularity distributions, and that perceived popularity can be shaped by artist- or track-level familiarity~\cite{DBLP:conf/chiir/FerwerdaIBS23}. This matters for user studies on calibration: if users evaluate unfamiliar tracks, their judgments may reflect lack of recognition rather than the popularity distribution of the list. In the present study, we therefore reduce the role of unfamiliarity by constructing recommendation lists from tracks users have recently consumed. This allows us to study perceived differences between popularity distributions under conditions where users are more likely to recognize the presented items.

Popularity calibration also implicitly treats a user's listening history as evidence of a popularity-related preference~\cite{DBLP:conf/um/AbdollahpouriMB21,DBLP:journals/tors/SilvaJ26}. However, in music recommendation, the relevant history is not obvious. A user's complete listening history may span many years, while recent listening may better reflect current taste. Moreover, not all tracks in a user's history may be available in the catalog used for computing popularity, and popularity estimates may differ depending on the dataset and time period considered~\cite{DBLP:conf/recsys/LesotaMRBKLS21,kowald_study_2023}. This raises the question of how robust popularity calibration is when the target distribution is estimated from incomplete or differently sized portions of user history.

Taken together, these issues underline the importance of studying popularity calibration not only as an offline metric, but also as a user-facing construct. Prior work has already started to compare computational and perceived popularity bias or miscalibration in music recommendation~\cite{DBLP:conf/chiir/FerwerdaIBS23,DBLP:conf/sigir/LesotaEDFKLRS23}. We build on this line of work by examining the effects of item familiarity and incomplete user history. If calibration is intended to represent a user-centered notion of recommendation quality, then users should perceive meaningful differences between differently calibrated lists. Furthermore, calibrated lists should provide measurable value to users, and the underlying metric should remain robust in the face of realistic limitations in user-history data.

In this work, we conduct a user study in which participants evaluate music track lists with different popularity compositions. Rather than relying on standard recommender algorithms, we use a controlled naive recommender that constructs lists from tracks in users' recent listening histories. This design allows us to compare lists dominated by high-popularity tracks, lists dominated by low-popularity tracks, and popularity-calibrated lists while reducing the confounding effect of unfamiliar items. We then relate participants' track-level and list-level judgments to offline popularity-calibration measurements, including Jensen--Shannon divergence (JSD) between the popularity distribution of the list and the user's estimated historical popularity profile. More precisely, we address the following research questions:

\begin{itemize}
\item \textbf{RQ1:} Do users perceive differences in popularity levels between recommendation lists generated with different calibration targets?

\item \textbf{RQ2:} Do users prefer popularity-calibrated recommendation lists over lists dominated by high- or low-popularity items?

\item \textbf{RQ3:} How robust is the common offline popularity-calibration measurement framework under different levels of item familiarity, list composition, and available user-history?

\item \textbf{RQ4:} How do computationally assigned popularity labels align with users' own popularity judgments?
\end{itemize}

By addressing these questions, this paper contributes to a more critical understanding of popularity calibration as both an offline measurement framework and a user-facing construct. We examine whether users perceive the intended differences between differently calibrated lists, whether calibrated lists are actually preferred, how sensitive calibration estimates are to incomplete user-history information, and whether computational popularity labels align with users' own judgments of item popularity.

\section{Related Work}

\subsection{Foundations and extensions of popularity calibration}

\citet{DBLP:conf/um/AbdollahpouriMB21} introduced popularity calibration as a user-centered debiasing method designed to align the distribution of head, mid, and tail items in personalized recommendation lists. In contrast to global popularity-bias measures, popularity calibration evaluates whether the popularity distribution of a recommendation list matches the popularity distribution of a user's past interactions. This makes calibration a user-centered metric: a list dominated by popular items may be calibrated for one user, but miscalibrated for another.

Subsequent research has applied and extended this framework across different domains and evaluation settings. In the movie streaming sector, \citet{DBLP:conf/bias/KlimashevskaiaE22} found that popularity calibration can positively affect consumer fairness, although its effectiveness for provider fairness is more limited. Within the music, movie, and anime domains, \citet{kowald_study_2023} demonstrated that calibration and miscalibration can be used as metrics to measure popularity bias in recommender systems. Related work has also used calibration-based metrics to examine whether popularity bias affects users of different genders differently in music recommendation~\citep{DBLP:conf/recsys/LesotaMRBKLS21}. More recently, \citet{forster2025exploring} combined popularity calibration with context-aware mechanisms to improve consumer-side fairness in point-of-interest recommendation.

Although popularity calibration is often evaluated through offline metrics, recent work has started to examine calibration and related bias-mitigation approaches in user-facing settings~\cite{DBLP:conf/recsys/KlimashevskaiaE23,DBLP:conf/recsys/UngruhDVPH24,DBLP:conf/um/AlvesJSM24}. This line of work is important because calibration can change the composition of recommendation lists in ways that are measurable computationally, but not necessarily noticed, understood, or valued by users. For example, \citet{DBLP:conf/um/AlvesJSM24} found that users do not necessarily perceive calibrated recommendations as fairer unless the system explicitly explains the purpose of calibration. These findings suggest that the computational value of calibration and its perceived value for users should be treated as related but distinct questions.

A second open issue concerns the robustness of the calibration estimate itself. Popularity calibration depends not only on the recommendation list, but also on how item popularity is computed, how popularity categories are defined, and how the user's historical profile is represented. This is particularly important when calibration is interpreted as a user-centered metric, because missing, unmatched, or temporally uneven interaction histories may change the target distribution against which recommendations are evaluated.

\subsection{Perceived vs. computational bias in recommendation}

A growing body of work has examined whether computational measures of popularity bias, fairness, or miscalibration correspond to users' perceptions. This distinction is particularly relevant for music recommender systems, where users may judge recommendations not only by their computed popularity distribution, but also through artist recognition, track familiarity, and perceived fit with their music taste.

Ferwerda et al.~\cite{DBLP:conf/chiir/FerwerdaIBS23} explored how users perceive popularity and fairness in music recommendations. Their findings suggest that users often do not recognize differences between presented popularity distributions, and that perceived popularity is more strongly associated with artists than with individual songs. This indicates that differences in computational popularity bias may not always be readily apparent to users.

Closely related to the present work, Lesota et al.~\cite{DBLP:conf/sigir/LesotaEDFKLRS23} compared computational and perceived popularity miscalibration in music recommendation. They showed that JSD, a computational metric for popularity miscalibration, can correlate with user perception of recommendation lists. However, item familiarity remained an important limitation: when users are unfamiliar with recommended items, their ability to judge popularity or miscalibration is constrained. These studies motivate a more detailed examination of the conditions under which computational and perceived popularity calibration align.

This creates a methodological tension for user studies. Unfamiliar recommendations may be closer to discovery-oriented recommendation scenarios, but they make it difficult to ask users to judge popularity, liking, or fit. Familiar items, in contrast, provide a cleaner setting for studying whether users can perceive differences in popularity composition, but may also increase the role of recognition and prior liking. The present work explicitly focuses on this tension by constructing lists from recently consumed tracks.

\subsection{User evaluation and awareness in music recommender systems}

Music recommendation is a central domain for studying popularity bias and calibration because recommendation lists can affect both listener experience and exposure for artists or tracks~\cite{DBLP:conf/ecir/KowaldSL20,DBLP:conf/recsys/LesotaMRBKLS21,DBLP:conf/sigir/LesotaEDFKLRS23}. Prior work has used calibration-based metrics to analyze popularity bias in music recommender systems~\cite{DBLP:conf/recsys/LesotaMRBKLS21,kowald_study_2023}. At the same time, music recommendation makes user-centered evaluation difficult: users may respond strongly to whether they recognize a track, whether they like the artist, or whether the list fits their taste~\cite{DBLP:conf/chiir/FerwerdaIBS23}. Therefore, a list that is computationally calibrated is not necessarily experienced as better by the user.

Complementary insights are provided by Dinnissen et al.~\cite{DBLP:conf/ht/dinnissen2025role}, who conducted a mixed-methods user study combining qualitative and quantitative methods, including think-aloud protocols, interviews, observed user behavior, and questionnaires. Their work highlights that users may reflect on popularity, diversity, and fairness in music recommendation, but that such reflection does not necessarily translate directly into different selection behavior. Dokoupil et al.~\cite{DBLP:conf/recsys/DokoupilBP25,Dokoupil2024} further examined user perceptions of diversity and system objectives, emphasizing the nuanced ways users interpret and respond to system-driven fairness and diversity cues.

Our work builds on these foundations by studying popularity calibration under two conditions that are central to music recommendation but remain methodologically challenging: item familiarity and incomplete user history. Specifically, we examine whether users perceive differences between lists with different popularity compositions, whether calibrated lists are preferred over high- or low-popularity lists, whether offline calibration estimates remain robust under different familiarity and history-availability conditions, and whether computational popularity labels align with users' own judgments of item popularity.

\section{Methods and materials}

We address the research questions by conducting a user study in which participants evaluated music track lists with respect to familiarity, liking, perceived popularity, and perceived fit with their taste. We then compare these user judgments with offline measurements of popularity calibration. The study setup is designed to control two issues that complicate user studies on popularity calibration: variation in item familiarity and incomplete availability of user-history information. In this section, we describe how popularity calibration is measured, how item popularity labels are assigned, how the track lists are generated, which questions were posed, and how the resulting data are analyzed.

\subsection{Popularity calibration measurement}

A recommender is considered popularity-calibrated when, for a given user, the popularity distribution of the recommendation list matches the popularity distribution of the user's consumption history. The degree of popularity calibration is commonly evaluated at the user level as the divergence between two discrete probability distributions and then aggregated across users.

Following previous work, we distinguish between three item popularity categories, $c \in \mathcal{C}$, where
\[
\mathcal{C} = \{\texttt{HighPop}, \texttt{MidPop}, \texttt{LowPop}\}.
\]
We denote by $H_u$ the three-bin popularity distribution of the consumption history of user $u$, and by $R_u$ the corresponding three-bin popularity distribution of the recommendation list shown to user $u$. We quantify popularity miscalibration using Jensen-Shannon Divergence (JSD), shown in Eq.~\ref{eq:jsd}, where $H_u(c)$ is the proportion of items from category $c$ in the user's history distribution, and $R_u(c)$ is the corresponding proportion in the recommendation list. Lower JSD values indicate stronger popularity calibration, whereas higher values indicate stronger popularity miscalibration. As usual, terms with zero probability are treated as contributing zero to the sum.

\begin{equation} \label{eq:jsd}
\begin{aligned}
    JSD(H_u, R_u) = \frac{1}{2}\sum_{c}{H_u(c) \cdot \log_2\frac{2H_u(c)}{H_u(c) + R_u(c)}} + \\
    \frac{1}{2}\sum_{c}{R_u(c) \cdot \log_2\frac{2R_u(c)}{H_u(c) + R_u(c)}}
\end{aligned}
\end{equation}

\subsection{Dataset and popularity labeling}

We conduct our study in the domain of music recommendation and use the LFM-2B dataset~\cite{DBLP:conf/chiir/SchedlBLPPR22} to assign popularity labels to music tracks. The dataset contains user listening events, or scrobbles, from Last.fm, spanning from February 2005 to March 2020. We process the data in several steps to obtain a collection of tracks with reliable popularity estimates.

First, we match tracks to Spotify URIs using relaxed artist and track-name matching. This step helps filter out listening events involving non-music content, such as podcasts and audiobooks, which may sometimes be tracked on Last.fm. Since relaxed matching can map remixes or less popular versions of a track to the source track, we apply a stricter filtering step later. Second, we assign raw popularity values to tracks based on the number of unique users who interacted with each track. Third, following previous work~\cite{DBLP:conf/um/AbdollahpouriMB21,DBLP:conf/recsys/LesotaBWWTS25}, we assign popularity labels by ranking tracks according to their popularity. The most popular tracks, which together account for 20\% of the cumulative user-track popularity mass, are labeled \texttt{HighPop}. Analogously, the least popular tracks that jointly account for 20\% of the total popularity mass are labeled \texttt{LowPop}, and the remaining tracks are labeled \texttt{MidPop}.

Finally, for the study itself, we only consider tracks whose artist name and title exactly match the metadata associated with the matched Spotify URI. This allows us to be more confident about the popularity labels of the tracks shown to participants, while still allowing the broader dataset to contribute to the overall popularity distribution. After this filtering procedure, the collection contains approximately 3.4M unique tracks.

\subsection{Naive recommender}

Previous work repeatedly relies on established recommender algorithms to provide participants with recommendation lists for evaluation~\cite{DBLP:conf/chiir/FerwerdaIBS23}; for example, \citet{DBLP:conf/sigir/LesotaEDFKLRS23} generated personalized recommendations for each participant using several recommender algorithms to analyze the relation between user-perceived and computational popularity miscalibration. While this approach is representative of realistic recommendation scenarios, it can introduce limitations to user studies. In particular, two challenges are relevant to the present work: limited variability between recommendation lists and varying levels of participants' familiarity with recommended items.

Traditional recommender algorithms often produce lists biased toward more popular items~\cite{DBLP:conf/recsys/LesotaMRBKLS21,DBLP:conf/recsys/LesotaBWMLRS22}. While the degree of bias varies between algorithms, some list types, such as lists dominated by niche items, may remain underrepresented. This can limit the range of popularity compositions available for user evaluation. A second challenge is that participants may be unfamiliar with recommended items because these items are drawn from the full catalog rather than from previously experienced tracks. This makes it difficult for participants to judge whether a track is popular, whether they like it, or whether it fits their taste. In some study designs, this issue is addressed by allowing participants to consume or preview the items, for example, by listening to audio snippets. However, this is not always desirable if the study focuses on participants' existing recognition and recollection of items, or if the researchers want to avoid introducing additional content-based effects or causing excessive participant fatigue.

In this study, we therefore use a controlled naive recommender. We use the term naive recommender to refer to a controlled list generator rather than a competitive recommendation algorithm. The goal is not to maximize recommendation accuracy, but to construct personalized lists with predefined popularity compositions while keeping item familiarity high. The naive recommender composes lists from tracks in the participant's recent listening history according to a predefined popularity distribution.

We consider three list types:

\begin{itemize}
    \item \modeldummylowpop --- a list of 10 random items from the participant's recent listening history containing as many \texttt{LowPop} items as possible. If the participant consumed fewer than 10 \texttt{LowPop} items, the remaining positions are filled with \texttt{MidPop} items. This list type represents the most niche part of the participant's recent listening history from the perspective of computationally assigned popularity labels.

    \item \modeldummyhighpop --- a list of 10 random items from the participant's recent listening history containing as many \texttt{HighPop} items as possible. If the participant consumed fewer than 10 \texttt{HighPop} items, the remaining positions are filled with \texttt{MidPop} items. This list type represents the most mainstream part of the participant's recent listening history from the perspective of computationally assigned popularity labels.

    \item \modeldummycali --- a list of 10 items representing the popularity distribution of the participant's recent listening history. First, the popularity distribution of the participant's recent history is computed based on computationally assigned popularity labels. Then, this distribution is approximated as a discrete 10-item list. The required number of items from each popularity category is sampled from the participant's recent listening history.
\end{itemize}

\subsection{Capturing participants' perception}

Each participant was presented with six personalized lists of music tracks: two \modeldummyhighpop lists, two \modeldummylowpop lists, and two \modeldummycali lists. The order of the six lists was randomized for each participant. Participants evaluated each list both at the list level and at the track level. The list-level questions are shown in Table~\ref{tab:list_q}, and the track-level questions are shown in Table~\ref{tab:track_q}.

At the track level, participants first indicated whether they knew each track. The liking and perceived-popularity questions were only shown for tracks that participants indicated as familiar. This design avoids asking participants to judge the popularity or appeal of tracks they did not recognize. For the analysis, track-level responses were aggregated into counts per list. In particular, $Know_{tr}$ denotes the number of known tracks in a list, $Like_{tr}$ denotes the number of familiar tracks marked as liked, $H.Pop._{tr}$ denotes the number of familiar tracks judged as closer to more popular on the spectrum of popularity of the participant's music taste, and $L.Pop._{tr}$ denotes the number of familiar tracks judged as closer to less popular among the participant's usually listened music.

\begin{table}[t]
    \centering
    \caption{List-level evaluation questions. Responses were given on a five-point Likert scale.}
    \label{tab:list_q}
    \begin{tabular}{ll}
    \toprule
        Code & Question \\
    \midrule
        $Know_{l}$ & ``The list contains tracks I know''\\
        $Like_{l}$ & ``I like this list of tracks'' \\
        $Pop._{l}$ & ``The list contains a lot of popular items''\\
        $Taste_{l}$ & ``The list represents my music taste well''\\
        $Unpop._{l}$ & ``The list contains a lot of unpopular items''\\
    \bottomrule
    \end{tabular}
\end{table}

\begin{table}[t]
    \centering
    \caption{Track-level evaluation questions and response options.}
    \label{tab:track_q}
    \begin{tabular}{lp{0.43\textwidth}p{0.35\textwidth}}
    \toprule
        Code & Statement & Response options \\
    \midrule
        $Know_{tr}$ & ``I know this track'' & ``no'' / ``yes''\\
        $Like_{tr}$ & ``I like this track'' & ``dislike'' / ``neutral'' / ``like''\\
        $Pop._{tr}$ & ``Among the music I listen to, this track is closer to'' & ``less popular'' / ``average'' / ``more popular''\\
    \bottomrule
    \end{tabular}
\end{table}

\subsection{Participants}

We recruited participants through Prolific\footnote{\url{https://www.prolific.com/}}. Eligible participants were identified through a pre-screening study as active Last.fm users. In total, 80 participants were recruited for the main study. At the time of participation, participants had to satisfy the following inclusion criteria:

\begin{itemize}
    \item being an active user of at least one music streaming platform (self-reported);
    \item residing in the United States;
    \item having recorded at least 60 listening events on Last.fm during the previous 30 days;
    \item during the previous seven weeks, having listened to at least 60 distinct tracks that could be matched to the dataset used for popularity labeling.
\end{itemize}

For each participant, we retrieved up to 4000 of their most recent Last.fm listening events through the Last.fm API\footnote{\url{https://www.last.fm/api}}, restricted to events no older than seven weeks before the start of the study. Because the study remained open for one week, this ensured that every track shown to a participant had been listened to by that participant at least once, no more than eight weeks before participation.

Participants had an estimated completion time of 15 minutes. The observed median completion time was 14.2 minutes. Following our exclusion criterion, we removed participants who completed the study in less than 10 minutes. This resulted in a final sample of 60 participants with an average age of 31.4 ($std = 9.5$) and the gender distribution over Diverse/Female/Male as 5/27/28. The \modeldummyhighpop lists created for the participants contain on average 9 \texttt{HighPop} items with median of 10. Out of 60 participants 11 had fewer than 10 \texttt{HighPop} items in their listening histories,  of which 2 consumed no \texttt{HighPop} items at all (therefore the two received \texttt{MidPop} items in their \modeldummyhighpop lists). All \modeldummylowpop lists created for the participants had all 10 items from the \texttt{LowPop} category.

\subsection{Ethical considerations}

The study did not undergo formal review by an institutional ethics board, as no applicable institutional ethics-review procedure was available for this type of study. We nevertheless followed standard ethical research practices and Prolific's requirements for participant recruitment, study description, informed participation, compensation, and withdrawal. Participants were informed about the purpose of the study, the type of data used, the expected duration of the task, and their right to discontinue participation without adverse effects. Participation was voluntary, and participants were compensated through Prolific.

Because the study relies on participants' recent Last.fm listening histories, we treated listening data as potentially identifying behavioral data. The data were used only to construct personalized study lists, assign popularity labels, and conduct the analyses reported in this paper. Participant records were pseudonymized for analysis, and results are reported only in aggregate. We do not disclose individual participants' listening histories, Last.fm profiles, or item-level responses in a form that would allow participants to be identified.

\subsection{Analysis workflow}

For each list, we compute i) track-level and ii) list-level perceptual indicators. Track-level indicators are aggregated as counts out of 10, including the number of tracks known by the participant, liked by the participant, judged as highly popular, and judged as less popular. List-level indicators are taken directly from the five-point Likert-scale questions.

Table~\ref{tab:mops} reports mean values by list type: \modeldummyhighpop ($High$), \modeldummycali ($Cali$), \modeldummylowpop ($Low$). The descriptive values are averaged over lists of each type. 

We use Friedman tests to test for overall differences between the three list types and Wilcoxon signed-rank tests for pairwise comparisons. We apply a Bonferroni-corrected significance threshold of $\alpha = 0.05/36$ across the tests reported in Table~\ref{tab:mops}.

To analyze the relation between computational popularity miscalibration and participants' perception, we compute repeated-measures correlations between JSD and list-level perceptions of popularity and unpopularity. We conduct these analyses under different thresholds for item familiarity and available user history. Specifically, we compare settings in which participants are familiar with at least five or at least eight tracks in each list, and settings in which the user's popularity profile is estimated from different minimum amounts of matched user history.

Finally, to compare computational popularity labels with participant-judged popularity labels, we compute agreement between the computational category of each track and the participant's track-level judgment. We summarize this agreement using Cohen's $\kappa$ and Kendall's $\tau$, and additionally inspect category-level agreement for \texttt{LowPop}, \texttt{MidPop}, and \texttt{HighPop} items.

\section{Results}

We organize the results according to the four research questions. First, we examine whether participants perceived differences between the three list types. Second, we analyze whether popularity-calibrated lists were preferred over lists dominated by high- or low-popularity tracks. Third, we investigate how the relation between computational popularity miscalibration and perceived list popularity changes under different familiarity and user-history conditions. Finally, we compare computationally assigned popularity labels with participants' own track-level popularity judgments.

\paragraph{Perceived differences between list types.}
Table~\ref{tab:mops} reports mean perceptual indicators for \modeldummyhighpop, \modeldummycali, and \modeldummylowpop lists. The values are averaged over lists of each type. The results show that participants perceived clear differences in popularity composition between the three list types.

\begin{table}[t]
    \centering
    \caption{Comparison of mean opinion scores over perceptual indicators. Values are averaged over lists of each type; Significant pairwise differences from \modeldummyhighpop ($High$), \modeldummycali ($Cali$), and \modeldummylowpop ($Low$) lists are marked as $^h$, $^c$, and $^l$, respectively (Wilcoxon signed-rank test). Friedman test statistics ($\chi^2(2)$) for each indicator are reported in the last row, with significant differences marked by $^*$ (Bonferroni-corrected $\alpha = 0.05/36$).}
    \label{tab:mops}
    \resizebox{\textwidth}{!}{
    \begin{tabular}{c||cccc|ccccc}
    \toprule
    List & \multicolumn{4}{c|}{Track count (out of 10)} & \multicolumn{5}{c}{Whole-list evaluation (Likert-5)} \\
    Type & $Like_{tr}$ & $H.Pop._{tr}$ & $L.Pop._{tr}$ & $Know_{tr}$ & $Like_{l}$ & $Pop._{l}$ & $Unpop._{l}$ & $Know_{l}$ & $Taste_{l}$ \\
    \midrule
    $High$ & $\textbf{7.96}^{l}$ & $\textbf{4.81}^{cl}$ & $1.40^{cl}$ & $\textbf{9.12}^{cl}$ & $\textbf{4.51}$ & $\textbf{4.15}^{cl}$ & $2.06^{cl}$ & $\textbf{4.87}^{l}$ & $3.83$ \\
    $Cali$ & $7.48$ & $2.89^{hl}$ & $2.60^{hl}$ & $8.57^{h}$ & $4.46$ & $3.32^{hl}$ & $2.87^{hl}$ & $4.71$ & $\textbf{3.87}$ \\
    $Low$ & $7.05^{h}$ & $1.59^{hc}$ & $\textbf{3.80}^{hc}$ & $8.47^{h}$ & $4.26$ & $2.41^{hc}$ & $\textbf{3.50}^{hc}$ & $4.62^{h}$ & $3.54$ \\
    \midrule
    Friedman $\chi^2(2)$ & $20.78^{*}$ & $89.75^{*}$ & $55.03^{*}$ & $16.17^{*}$ & $8.51$ & $90.60^{*}$ & $71.38^{*}$ & $14.19^{*}$ & $4.98$ \\
    \bottomrule
    \end{tabular}
    }
\end{table}

At the track level, \modeldummyhighpop lists contained the highest number of tracks judged as highly popular ($H.Pop._{tr}=4.81$), followed by \modeldummycali lists ($2.89$) and \modeldummylowpop lists ($1.59$). The reverse pattern appears for tracks judged as less popular: \modeldummylowpop lists received the highest number of low-popularity judgments ($L.Pop._{tr}=3.80$), followed by \modeldummycali lists ($2.60$) and \modeldummyhighpop lists ($1.40$). These differences are significant across all pairwise comparisons.

The same pattern is visible at the list level. Participants rated \modeldummyhighpop lists as containing the most popular items ($Pop._l=4.15$), followed by \modeldummycali lists ($3.32$) and \modeldummylowpop lists ($2.41$). Conversely, \modeldummylowpop lists were rated as containing the most unpopular items ($Unpop._l=3.50$), followed by \modeldummycali lists ($2.87$) and \modeldummyhighpop lists ($2.06$). The significant Friedman tests and pairwise Wilcoxon comparisons indicate that the list-generation procedure successfully produced lists that were perceived as different in their popularity composition.

It is also worth noting that familiarity was high across all conditions. On average, participants reported knowing more than eight tracks per list in all three conditions. However, \modeldummyhighpop lists were still perceived as slightly more familiar than the other two list types, both at the track level ($Know_{tr}=9.12$) and at the list level ($Know_l=4.87$). This suggests that constructing lists from recent listening histories reduced unfamiliarity substantially, but did not fully eliminate familiarity differences between popularity conditions.

\paragraph{Preference for calibrated lists.}
Although participants perceived clear popularity differences between list types, these differences did not translate into a clear preference for calibrated lists. At the list level, liking was high for all three list types: \modeldummyhighpop ($Like_l=4.51$), \modeldummycali ($4.46$), and \modeldummylowpop ($4.26$). The Friedman test for list liking was not significant after correction. Similarly, perceived taste fit did not differ significantly between the list types, with \modeldummycali lists showing only a small numerical advantage ($Taste_l=3.87$) over \modeldummyhighpop ($3.83$) and \modeldummylowpop ($3.54$) lists.

At the track level, participants liked more tracks in \modeldummyhighpop lists ($Like_{tr}=7.96$) than in \modeldummylowpop lists ($7.05$), while \modeldummycali lists fell between them ($7.48$). This suggests that popularity-calibrated lists were not clearly preferred over popularity-miscalibrated lists. Instead, when all lists were constructed from recently consumed and therefore familiar tracks, participants generally liked the lists regardless of their calibration condition. This result qualifies the user-centered interpretation of popularity calibration: users can perceive differences in popularity composition, but this does not necessarily mean that they prefer the calibrated condition.

\paragraph{Relation between JSD and perceived popularity.}
Tables~\ref{tab:rmcorr_all_5} and~\ref{tab:rmcorr_all_8} report repeated-measures correlations between JSD and list-level perceptions of popularity and unpopularity when all three list types are included.

\begin{table}
    \centering
    \caption{Repeated measures correlation with popularity miscalibration (JSD).
      Every participant is familiar with at least 5 tracks in each list. All three types of lists are included.}
      \label{tab:rmcorr_all_5}
    \begin{tabular}{ccc||cc|cc}
    \toprule
History matched & \# participants & \# lists & $r_{rm}(Pop._{l})$ & $p$ & $r_{rm}(Unpop._{l})$ & $p$ \\
\midrule
$60+$ & $60$ & $345$ & $0.151$ & $0.010$ & $-0.168$ & $0.004$ \\
$90+$ & $49$ & $283$ & $0.185$ & $0.004$ & $-0.184$ & $0.005$ \\
$120+$ & $41$ & $235$ & $0.186$ & $0.009$ & $-0.192$ & $0.007$ \\
\bottomrule
      \end{tabular}
\end{table}

\begin{table}
    \centering
    \caption{Repeated measures correlation with popularity miscalibration (JSD).
      Every participant is familiar with at least 8 tracks in each list. All three types of lists are included.}
      \label{tab:rmcorr_all_8}
    \begin{tabular}{ccc||cc|cc}
    \toprule
History matched & \# participants & \# lists & $r_{rm}(Pop._{l})$ & $p$ & $r_{rm}(Unpop._{l})$ & $p$ \\
\midrule
$60+$ & $56$ & $253$ & $0.218$ & $0.002$ & $-0.194$ & $0.006$ \\
$90+$ & $45$ & $199$ & $0.278$ & $<.001$ & $-0.189$ & $0.019$ \\
$120+$ & $37$ & $162$ & $0.28$ & $0.001$ & $-0.211$ & $0.018$ \\
\bottomrule
      \end{tabular}
\end{table}

When participants were familiar with at least five tracks in each list, the correlations between JSD and perceived list popularity were positive but weak, ranging from $r=.151$ with at least 60 matched history items to $r=.186$ with at least 120 matched history items. The correlations between JSD and perceived list unpopularity were negative and similarly weak, ranging from $r=-.168$ to $r=-.192$.

When the familiarity threshold was increased to at least eight known tracks per list, the relation between JSD and perceived popularity became somewhat stronger. Correlations with perceived list popularity increased from $r=.218$ to $r=.280$ across the three history thresholds. Correlations with perceived list unpopularity remained negative and modest, ranging from $r=-.194$ to $r=-.211$. These results suggest that the relation between computational miscalibration and perceived popularity becomes clearer when participants are more familiar with the evaluated tracks. However, when all three list types are considered together, the overall correlations remain modest.

\paragraph{Effect of list composition on JSD correlations.}
A clearer pattern appears when the analysis is restricted to \modeldummyhighpop and \modeldummycali lists.

\begin{table}
    \centering
    \caption{Repeated measures correlation with popularity miscalibration (JSD).
      Every participant is familiar with at least 5 tracks in each list. Only \modeldummyhighpop and \modeldummycali lists are included.}
      \label{tab:rmcorr_highcali_5}
    \begin{tabular}{ccc||cc|cc}
    \toprule
History matched & \# participants & \# lists & $r_{rm}(Pop._{l})$ & $p$ & $r_{rm}(Unpop._{l})$ & $p$ \\
\midrule
$60+$ & $60$ & $235$ & $0.526$ & $<.001$ & $-0.495$ & $<.001$ \\
$90+$ & $49$ & $193$ & $0.540$ & $<.001$ & $-0.493$ & $<.001$ \\
$120+$ & $41$ & $161$ & $0.550$ & $<.001$ & $-0.478$ & $<.001$ \\
\bottomrule
      \end{tabular}
\end{table}
%%%

\begin{table}
    \centering
    \caption{Repeated measures correlation with popularity miscalibration (JSD).
      Every participant is familiar with at least 8 tracks in each list. Only \modeldummyhighpop and \modeldummycali lists are included.}
      \label{tab:rmcorr_highcali_8}
    \begin{tabular}{ccc||cc|cc}
    \toprule
History matched & \# participants & \# lists & $r_{rm}(Pop._{l})$ & $p$ & $r_{rm}(Unpop._{l})$ & $p$ \\
\midrule
$60+$ & $52$ & $173$ & $0.511$ & $<.001$ & $-0.469$ & $<.001$ \\
$90+$ & $41$ & $135$ & $0.537$ & $<.001$ & $-0.438$ & $<.001$ \\
$120+$ & $33$ & $111$ & $0.523$ & $<.001$ & $-0.402$ & $<.001$ \\
\bottomrule
      \end{tabular}
\end{table}

As shown in Tables~\ref{tab:rmcorr_highcali_5} and~\ref{tab:rmcorr_highcali_8}, correlations between JSD and perceived list popularity are substantially stronger in this setting. With a familiarity threshold of at least five known tracks per list, correlations range from $r=.526$ to $r=.550$. With a stricter threshold of at least eight known tracks, correlations remain similarly strong, ranging from $r=.511$ to $r=.537$. The corresponding correlations with perceived list unpopularity are negative and also substantial.

This indicates that JSD aligns more clearly with user perception when the comparison is between mainstream-heavy and calibrated lists. When \modeldummylowpop lists are included, the relation becomes weaker. One likely reason is that JSD measures the magnitude of divergence between the list and the user's historical popularity profile, but does not by itself encode whether the divergence comes from a shift toward more popular or less popular items. Thus, JSD can indicate that a list is miscalibrated, but additional information is needed to interpret the direction of the miscalibration.

\paragraph{Influence of available user history.}
The history-threshold comparisons show that the relation between JSD and perceived popularity is sensitive to how much matched user history is available. In the all-list setting, increasing the minimum amount of matched history from 60 to 90 or 120 tracks generally increases the correlation between JSD and perceived popularity, although the effect remains modest. This suggests that popularity calibration estimates become somewhat more aligned with user perception when the user's target popularity profile is based on more matched listening history.

However, the pattern is not uniformly stronger across all settings. In the \modeldummyhighpop--\modeldummycali comparison, correlations are already strong at the 60-track threshold and change only slightly as the amount of matched history increases. This suggests that the effect of available user history depends on list composition. Additional matched history may matter most when the analysis includes multiple forms of miscalibration, whereas the distinction between high-popularity-heavy and calibrated lists is already captured relatively clearly with less history.

\paragraph{Alignment between computational and participant-judged popularity labels.}
Finally, we compare computational popularity labels with participants' track-level popularity judgments. Agreement between the two categorizations was weak overall, with a mean Cohen's $\kappa$ of $.21$ and a mean Kendall's $\tau$ of $.36$. Agreement was higher for the extreme categories than for the middle category: $62\%$ of participant-assigned low-popularity judgments matched the corresponding computational label, and $58\%$ of high-popularity judgments matched the computational label. In contrast, only $24\%$ of participant-assigned mid-popularity judgments matched the computational mid-popularity category.

This weak alignment indicates that computational popularity labels and participant-judged popularity labels should not be treated as interchangeable. Participants appear to identify relatively popular and relatively unpopular tracks more consistently than mid-popularity tracks, but their judgments do not fully correspond to the popularity bins used for offline calibration. This complicates the interpretation of popularity calibration as a user-centered metric, because the target distribution is based on computational categories that only partially align with users' own popularity perceptions.

\paragraph{Summary.}
Overall, the results show that participants perceive the intended differences between \modeldummyhighpop, \modeldummycali, and \modeldummylowpop lists. However, calibrated lists are not clearly preferred over high- or low-popularity-heavy lists. The relation between JSD and perceived popularity is present but sensitive to item familiarity, list composition, and the amount of matched user history. Finally, the weak agreement between computational and user-judged popularity labels raises questions about how directly offline popularity categories can represent users' own perceptions of popularity.

\section{Discussion}

Our results provide a nuanced view of popularity calibration as a user-centered metric in music recommendation. On the one hand, participants clearly perceived the intended differences between lists with different popularity compositions. Lists constructed to be high-popularity-heavy were judged as more popular, while low-popularity-heavy lists were judged as more unpopular. This suggests that users can perceive differences in popularity composition when the presented tracks are sufficiently familiar. On the other hand, these perceived differences did not translate into a clear preference for popularity-calibrated lists. This distinction is important: a calibration metric may capture a meaningful computational property of a recommendation list, but this does not automatically mean that users experience calibrated lists as more useful, more enjoyable, or more representative of their taste.

\subsection{Popularity calibration as a user-centered metric}

Popularity calibration is often motivated as a user-centered alternative to global popularity-bias measures~\cite{DBLP:conf/um/AbdollahpouriMB21,DBLP:journals/tors/SilvaJ26}. Rather than asking whether a system recommends too many popular items overall, calibration asks whether the popularity profile of a recommendation list matches the user's historical consumption profile. This makes the metric attractive because it recognizes that popularity preference may differ between users. A mainstream-oriented listener may be expected to receive more popular tracks, while a listener with a stronger long-tail profile may reasonably receive more niche tracks.

However, our results show that this user-centered framing should be interpreted carefully. Participants perceived differences between high-popularity-heavy, low-popularity-heavy, and calibrated lists, but they did not clearly prefer the calibrated lists. List-level liking was high across all three conditions, and perceived taste fit did not significantly differ between list types. This suggests that calibration may describe a certain kind of alignment with past behavior, but not necessarily the kind of alignment users value most when evaluating music recommendations.

One possible explanation is that our study deliberately controlled item familiarity by constructing lists from recently consumed tracks. This design makes it easier to study whether users perceive differences in popularity composition, but it also means that all lists contain relatively familiar and recently relevant items. Under these conditions, even a popularity-miscalibrated list may still be liked because it consists of tracks the user knows and has listened to recently. This interpretation aligns with prior work suggesting that perceived popularity and user evaluations of music recommendations can be strongly affected by familiarity and recognition~\cite{DBLP:conf/chiir/FerwerdaIBS23,DBLP:conf/ht/dinnissen2025role}.

\subsection{What should be the target of calibration?}

A central question raised by our findings is what should count as the target distribution for popularity calibration. In standard offline evaluation, the target is usually derived from the user's historical interactions~\cite{DBLP:conf/um/AbdollahpouriMB21,DBLP:journals/tors/SilvaJ26}. However, user histories are not neutral representations of preference. They depend on the time span covered by the dataset, the availability of item metadata, the platform from which interactions are collected, and the filtering decisions made during preprocessing.

This is particularly important in music recommendation. A user's complete listening history may span many years, but recent listening may better represent current taste. Conversely, a short recent history may be too narrow to capture stable preference patterns. Prior work on music preferences and mainstreaminess already shows that music consumption can vary across users, countries, and temporal contexts~\cite{BauerSchedlMainstream,DBLP:journals/tismir/LexKS20,DBLP:conf/ismir/LesotaPBLRS22}. The issue becomes even more complex when parts of the listening history cannot be matched to the item catalog used for popularity estimation. In such cases, the target popularity profile is not simply ``the user's history,'' but the subset of the user's history that remains observable after matching, filtering, and popularity labeling.

Our correlation analyses show that the relation between JSD and perceived list popularity changes with the amount of matched user history. This indicates that popularity calibration estimates are sensitive to how the user's historical profile is represented. As a result, calibration should not be treated as a purely objective property of a recommendation list. It is also a property of the data representation used to construct the user's target profile.

This has implications for offline evaluation. If different datasets cover different temporal spans, or if different users have histories of very different lengths, then the meaning of calibration may vary across users and domains. A ``complete history'' may represent years of listening for one user and weeks of activity for another. Therefore, future work should be more explicit about what portion of user history is treated as representative of popularity preference, and why.

\subsection{What does ``popular'' mean to users?}

Our results also show weak agreement between computational popularity labels and participant-judged popularity labels. This raises a second construct-validity issue: offline metrics usually define popularity through aggregate interaction data, whereas users may judge popularity through different cues. For example, a participant may judge a track as popular because they recognize the artist, because the song is widely known in their social context, or because it belongs to a mainstream genre. Conversely, a track may have many interactions in the dataset but still not be perceived as popular by a specific listener.

This concern is consistent with prior work showing that perceived popularity in music recommendation may be associated more strongly with artists than with individual songs~\cite{DBLP:conf/chiir/FerwerdaIBS23}. It also connects to broader work on music mainstreaminess, where popularity can be defined at different levels, such as artist playcounts, listener counts, global mainstreaminess, or country-specific mainstreaminess~\cite{BauerSchedlMainstream}. In our study, computational labels and user judgments aligned only weakly overall, with stronger agreement for the extreme categories than for the middle category. This suggests that users may be relatively able to identify very popular or very niche tracks, while mid-popularity items are harder to classify consistently.

For popularity calibration, this matters because the metric depends on computational popularity bins. If these bins only partially align with users' own judgments, then a list that is computationally calibrated may not be perceived as calibrated by the user. This does not invalidate computational calibration, but it clarifies what it measures: alignment with a dataset-derived popularity distribution, not necessarily alignment with the user's subjective understanding of popularity.

\subsection{Direction of miscalibration}

Another implication concerns the direction of miscalibration. JSD measures the divergence between the user's historical popularity distribution and the recommendation-list distribution~\cite{DBLP:journals/tit/Lin91}. It captures the magnitude of miscalibration, but not whether the list shifts the user toward more popular or less popular items. Our results show that JSD correlates more strongly with perceived popularity when the comparison is between high-popularity-heavy and calibrated lists. When low-popularity-heavy lists are included, the correlations become weaker.

This suggests that the interpretation of JSD depends on list composition. A high JSD value can indicate that a list is miscalibrated, but it does not explain how it is miscalibrated. For user-centered evaluation, this distinction is important. A list that is too mainstream and a list that is too niche may have similar JSD values, but they may be perceived very differently by users and may have different implications for personalization, discovery, and fairness.

Future work should therefore consider complementing symmetric divergence-based calibration metrics with directional indicators. For example, reporting whether a list shifts the user toward more high-popularity or low-popularity items would make calibration results easier to interpret. This would also help distinguish between calibration as a personalization objective and calibration as a fairness or exposure-related objective.

\subsection{Calibration, fairness, and user value}

Popularity calibration is often connected to fairness because reducing popularity bias may improve exposure for less popular items or providers~\cite{DBLP:conf/um/AbdollahpouriMB21,DBLP:conf/recsys/AbdollahpouriMB20,DBLP:conf/recsys/LesotaMRBKLS21}. However, our results underline that fairness value and user-perceived value are not the same. A calibrated list may be desirable from the perspective of reducing popularity distortion, but users may not necessarily prefer it or perceive it as a better fit.

This distinction is consistent with recent work showing that users do not always perceive calibrated or fairness-oriented recommendation interventions in the way system designers intend~\cite{DBLP:conf/um/AlvesJSM24,DBLP:conf/recsys/UngruhDVPH24,DBLP:conf/recsys/KlimashevskaiaE23}. It also connects to broader work on fairness in recommender systems, which emphasizes that fairness is a multi-stakeholder and context-dependent concept~\cite{DBLP:journals/ftir/Ekstrand0B022,DBLP:journals/umuai/DeldjooJBDZ24,DBLP:conf/um/SchedlRLGG22}.

This means that the purpose of calibration should be made explicit. If the goal is personalization, then calibration should be evaluated in terms of whether it improves perceived taste fit, satisfaction, or long-term usefulness. If the goal is fairness or exposure diversity, then calibration may be valuable even when users do not immediately prefer calibrated lists. These are different normative goals, and the same metric should not be assumed to satisfy all of them equally.

\subsection{Limitations}

Several limitations should be considered when interpreting our findings. First, our naive recommender was designed to control list composition and item familiarity rather than to compete with state-of-the-art recommender algorithms. This design is useful for isolating the role of popularity composition, but it does not fully reflect how users encounter new recommendations in real systems.

Second, because lists were constructed from recently consumed tracks, familiarity was high across all conditions. This was intentional, but it may also reduce the observable value of calibration. Users may like all lists because the items are familiar and recently relevant, regardless of whether the list is calibrated. Future studies could systematically vary familiarity to examine when calibration becomes more or less valuable.

Third, our computational popularity labels are derived from the available dataset and preprocessing pipeline. Different datasets, time windows, or popularity definitions could produce different labels and, therefore, different calibration estimates. This is not only a limitation of the present study but also an important methodological issue for popularity calibration research more broadly.

Finally, our study focuses on perceived popularity, liking, familiarity, and taste fit. Other outcomes, such as perceived fairness, discovery value, novelty, long-term satisfaction, or artist-level exposure, may reveal additional aspects of the value of calibration. Prior work on diversity, nudging, and user perceptions of recommendation objectives suggests that such outcomes may be important for understanding the broader user experience of music recommender systems~\cite{Dokoupil2024,DBLP:conf/recsys/DokoupilBP25,DBLP:conf/rs/liang2022exploring,porcaro2024assessing}.

\subsection{Implications}

The main implication of our study is that popularity calibration should not be treated as self-evidently user-centered simply because it uses the user's historical profile as a reference point. The metric depends on how that profile is constructed, how popularity is defined, and whether computational popularity categories correspond to users' own judgments. Our findings suggest that users can perceive differences in popularity composition, but that calibrated lists are not necessarily preferred, and that calibration estimates are sensitive to familiarity, list composition, and available user history.

For researchers, this means that popularity calibration should be reported with greater attention to measurement choices: the time span of user history, the amount of matched history, the popularity-binning strategy, and whether miscalibration is driven by shifts toward more popular or less popular items. For system designers, the findings suggest that calibration may be useful as one component of recommendation evaluation, but should be interpreted alongside user-facing measures such as liking, familiarity, perceived taste fit, and perceived discovery value.

\section{Conclusion}

Popularity calibration is commonly treated as a user-centered way to evaluate and mitigate popularity bias in recommender systems. In this paper, we examined this assumption in the context of music recommendation, focusing on item familiarity, list composition, and incomplete user-history information. We conducted a user study in which participants evaluated track lists with different popularity compositions, generated from their recent listening histories, and compared their judgments with offline popularity-calibration estimates.

With respect to \textbf{RQ1}, our results show that participants perceived clear differences between lists generated with different calibration targets. High-popularity-heavy lists were perceived as more popular, low-popularity-heavy lists were perceived as more unpopular, and calibrated lists generally fell between these two conditions. This indicates that users can perceive differences in popularity composition when the recommended items are sufficiently familiar.

For \textbf{RQ2}, however, these perceptual differences did not translate into a clear preference for calibrated lists. List liking and perceived taste did not significantly differ between the list types. This suggests that popularity calibration captures a meaningful property of recommendation lists, but that this property should not be equated directly with user appreciation or perceived preference.

Regarding \textbf{RQ3}, we found that the relation between JSD-based popularity miscalibration and perceived list popularity depends on familiarity, list composition, and the amount of available matched user history. JSD aligned more strongly with perceived popularity when comparing high-popularity-heavy and calibrated lists, but the relation became weaker when low-popularity-heavy lists were included. This highlights that offline calibration estimates are sensitive to both the available user-history information and the direction of miscalibration.

Finally, for \textbf{RQ4}, the agreement between computational popularity labels and participant-judged popularity labels was weak ($\kappa=.21$, $\tau=.36$). This suggests that dataset-derived popularity categories only partially capture users' perception of item popularity. This has implications for the use of popularity calibration as a user-centered metric, because the metric depends on computational labels that may not fully align with users' subjective popularity judgments.

Overall, our findings call for a more careful interpretation of popularity calibration in music recommender systems. Calibration remains useful as an offline measurement framework, but its user-centered value depends on how popularity is defined, how user history is represented, and whether calibrated lists are actually perceived and valued by users. Future work should further investigate directional calibration measures, alternative representations of user-history targets, and user-facing outcomes beyond liking, such as discovery value, perceived fairness, and long-term satisfaction.

\bibliographystyle{ACM-Reference-Format}

%DBLP formatted references
\bibliography{references_streamlined}

@inproceedings{DBLP:conf/sigir/LesotaEDFKLRS23,
  author       = {Oleg Lesota and
                  Gustavo Escobedo and
                  Yashar Deldjoo and
                  Bruce Ferwerda and
                  Simone Kopeinik and
                  Elisabeth Lex and
                  Navid Rekabsaz and
                  Markus Schedl},
  editor       = {Hsin{-}Hsi Chen and
                  Wei{-}Jou (Edward) Duh and
                  Hen{-}Hsen Huang and
                  Makoto P. Kato and
                  Josiane Mothe and
                  Barbara Poblete},
  title        = {Computational Versus Perceived Popularity Miscalibration in Recommender
                  Systems},
  booktitle    = {Proceedings of the 46th International {ACM} {SIGIR} Conference on
                  Research and Development in Information Retrieval, {SIGIR} 2023, Taipei,
                  Taiwan, July 23-27, 2023},
  pages        = {1889--1893},
  publisher    = {{ACM}},
  year         = {2023},
  url          = {https://doi.org/10.1145/3539618.3591964},
  doi          = {10.1145/3539618.3591964},
  bibsource    = {dblp computer science bibliography, https://dblp.org}
}

@inproceedings{DBLP:conf/ismir/LesotaPBLRS22,
  author       = {Oleg Lesota and
                  Emilia Parada{-}Cabaleiro and
                  Stefan Brandl and
                  Elisabeth Lex and
                  Navid Rekabsaz and
                  Markus Schedl},
  editor       = {Preeti Rao and
                  Hema A. Murthy and
                  Ajay Srinivasamurthy and
                  Rachel M. Bittner and
                  Rafael Caro Repetto and
                  Masataka Goto and
                  Xavier Serra and
                  Marius Miron},
  title        = {Traces of Globalization in Online Music Consumption Patterns and Results
                  of Recommendation Algorithms},
  booktitle    = {Proceedings of the 23rd International Society for Music Information
                  Retrieval Conference, {ISMIR} 2022, Bengaluru, India, December 4-8,
                  2022},
  pages        = {291--297},
  year         = {2022},
  url          = {https://archives.ismir.net/ismir2022/paper/000034.pdf},
  bibsource    = {dblp computer science bibliography, https://dblp.org}
}

@article{DBLP:journals/tismir/LexKS20,
  author       = {Elisabeth Lex and
                  Dominik Kowald and
                  Markus Schedl},
  title        = {Modeling Popularity and Temporal Drift of Music Genre Preferences},
  journal      = {Trans. Int. Soc. Music. Inf. Retr.},
  volume       = {3},
  number       = {1},
  pages        = {17--30},
  year         = {2020},
  url          = {https://doi.org/10.5334/tismir.39},
  doi          = {10.5334/TISMIR.39},
  bibsource    = {dblp computer science bibliography, https://dblp.org}
}

@inproceedings{DBLP:conf/ecir/KowaldSL20,
  author    = {Dominik Kowald and
               Markus Schedl and
               Elisabeth Lex},
  title     = {The Unfairness of Popularity Bias in Music Recommendation: {A} Reproducibility
               Study},
  booktitle = {Advances in Information Retrieval - 42nd European Conference on {IR}
               Research, {ECIR} 2020, Lisbon, Portugal, April 14-17, 2020, Proceedings,
               Part {II}},
  series    = {Lecture Notes in Computer Science},
  volume    = {12036},
  pages     = {35--42},
  publisher = {Springer},
  year      = {2020},
  url       = {https://doi.org/10.1007/978-3-030-45442-5\_5},
  doi       = {10.1007/978-3-030-45442-5\_5},
  bibsource = {dblp computer science bibliography, https://dblp.org}
}

@inproceedings{DBLP:conf/um/AbdollahpouriMB21,
  author    = {Himan Abdollahpouri and
               Masoud Mansoury and
               Robin Burke and
               Bamshad Mobasher and
               Edward C. Malthouse},
  editor    = {Judith Masthoff and
               Eelco Herder and
               Nava Tintarev and
               Marko Tkalcic},
  title     = {User-centered Evaluation of Popularity Bias in Recommender Systems},
  booktitle = {Proceedings of the 29th {ACM} Conference on User Modeling, Adaptation
               and Personalization, {UMAP} 2021, Utrecht, The Netherlands, June,
               21-25, 2021},
  pages     = {119--129},
  publisher = {{ACM}},
  year      = {2021},
  url       = {https://doi.org/10.1145/3450613.3456821},
  doi       = {10.1145/3450613.3456821},
  bibsource = {dblp computer science bibliography, https://dblp.org}
}

@inproceedings{DBLP:conf/bias/KlimashevskaiaE22,
  author       = {Anastasiia Klimashevskaia and
                  Mehdi Elahi and
                  Dietmar Jannach and
                  Christoph Trattner and
                  Lars Skj{\ae}rven},
  editor       = {Ludovico Boratto and
                  Stefano Faralli and
                  Mirko Marras and
                  Giovanni Stilo},
  title        = {Mitigating Popularity Bias in Recommendation: Potential and Limits
                  of Calibration Approaches},
  booktitle    = {Advances in Bias and Fairness in Information Retrieval - Third International
                  Workshop, {BIAS} 2022, Stavanger, Norway, April 10, 2022, Revised
                  Selected Papers},
  series       = {Communications in Computer and Information Science},
  volume       = {1610},
  pages        = {82--90},
  publisher    = {Springer},
  year         = {2022},
  url          = {https://doi.org/10.1007/978-3-031-09316-6\_8},
  doi          = {10.1007/978-3-031-09316-6\_8},
  bibsource    = {dblp computer science bibliography, https://dblp.org}
}

@inproceedings{DBLP:conf/recsys/LesotaMRBKLS21,
  author       = {Oleg Lesota and
                  Alessandro B. Melchiorre and
                  Navid Rekabsaz and
                  Stefan Brandl and
                  Dominik Kowald and
                  Elisabeth Lex and
                  Markus Schedl},
  editor       = {Humberto Jes{\'{u}}s Corona Pamp{\'{\i}}n and
                  Martha A. Larson and
                  Martijn C. Willemsen and
                  Joseph A. Konstan and
                  Julian J. McAuley and
                  Jean Garcia{-}Gathright and
                  Bouke Huurnink and
                  Even Oldridge},
  title        = {Analyzing Item Popularity Bias of Music Recommender Systems: Are Different
                  Genders Equally Affected?},
  booktitle    = {RecSys '21: Fifteenth {ACM} Conference on Recommender Systems, Amsterdam,
                  The Netherlands, 27 September 2021 - 1 October 2021},
  pages        = {601--606},
  publisher    = {{ACM}},
  year         = {2021},
  url          = {https://doi.org/10.1145/3460231.3478843},
  doi          = {10.1145/3460231.3478843},
  bibsource    = {dblp computer science bibliography, https://dblp.org}
}

@inproceedings{forster2025exploring,
  title={Exploring the effect of context-awareness and popularity calibration on popularity bias in poi recommendations},
  author={Forster, Andrea and Kopeinik, Simone and Helic, Denis and Thalmann, Stefan and Kowald, Dominik},
  booktitle={Proceedings of the Nineteenth ACM Conference on Recommender Systems},
  pages={593--598},
  year={2025}
}

@article{DBLP:journals/tit/Lin91,
  author       = {Jianhua Lin},
  title        = {Divergence measures based on the Shannon entropy},
  journal      = {{IEEE} Trans. Inf. Theory},
  volume       = {37},
  number       = {1},
  pages        = {145--151},
  year         = {1991},
  url          = {https://doi.org/10.1109/18.61115},
  doi          = {10.1109/18.61115},
  bibsource    = {dblp computer science bibliography, https://dblp.org}
}

@inproceedings{DBLP:conf/chiir/SchedlBLPPR22,
  author       = {Markus Schedl and
                  Stefan Brandl and
                  Oleg Lesota and
                  Emilia Parada{-}Cabaleiro and
                  David Penz and
                  Navid Rekabsaz},
  editor       = {David Elsweiler},
  title        = {LFM-2b: {A} Dataset of Enriched Music Listening Events for Recommender
                  Systems Research and Fairness Analysis},
  booktitle    = {{CHIIR} '22: {ACM} {SIGIR} Conference on Human Information Interaction
                  and Retrieval, Regensburg, Germany, March 14 - 18, 2022},
  pages        = {337--341},
  publisher    = {{ACM}},
  year         = {2022},
  url          = {https://doi.org/10.1145/3498366.3505791},
  doi          = {10.1145/3498366.3505791},
  bibsource    = {dblp computer science bibliography, https://dblp.org}
}

@inproceedings{kowald_study_2023,
  title={A study on accuracy, miscalibration, and popularity bias in recommendations},
  author={Kowald, Dominik and Mayr, Gregor and Schedl, Markus and Lex, Elisabeth},
  booktitle={International Workshop on Algorithmic Bias in Search and Recommendation},
  pages={1--16},
  year={2023},
  organization={Springer}
}

@inproceedings{DBLP:conf/recsys/AbdollahpouriMB20,
  author       = {Himan Abdollahpouri and
                  Masoud Mansoury and
                  Robin Burke and
                  Bamshad Mobasher},
  editor       = {Rodrygo L. T. Santos and
                  Leandro Balby Marinho and
                  Elizabeth M. Daly and
                  Li Chen and
                  Kim Falk and
                  Noam Koenigstein and
                  Edleno Silva de Moura},
  title        = {The Connection Between Popularity Bias, Calibration, and Fairness
                  in Recommendation},
  booktitle    = {RecSys 2020: Fourteenth {ACM} Conference on Recommender Systems, Virtual
                  Event, Brazil, September 22-26, 2020},
  pages        = {726--731},
  publisher    = {{ACM}},
  year         = {2020},
  url          = {https://doi.org/10.1145/3383313.3418487},
  doi          = {10.1145/3383313.3418487},
  bibsource    = {dblp computer science bibliography, https://dblp.org}
}

@inproceedings{DBLP:conf/recsys/KlimashevskaiaE23,
  author       = {Anastasiia Klimashevskaia and
                  Mehdi Elahi and
                  Dietmar Jannach and
                  Lars Skj{\ae}rven and
                  Astrid Tessem and
                  Christoph Trattner},
  editor       = {Jie Zhang and
                  Li Chen and
                  Shlomo Berkovsky and
                  Min Zhang and
                  Tommaso Di Noia and
                  Justin Basilico and
                  Luiz Pizzato and
                  Yang Song},
  title        = {Evaluating The Effects of Calibrated Popularity Bias Mitigation: {A}
                  Field Study},
  booktitle    = {Proceedings of the 17th {ACM} Conference on Recommender Systems, RecSys
                  2023, Singapore, Singapore, September 18-22, 2023},
  pages        = {1084--1089},
  publisher    = {{ACM}},
  year         = {2023},
  url          = {https://doi.org/10.1145/3604915.3610637},
  doi          = {10.1145/3604915.3610637},
  bibsource    = {dblp computer science bibliography, https://dblp.org}
}

@inproceedings{DBLP:conf/recsys/LesotaBWMLRS22,
  author       = {Oleg Lesota and
                  Stefan Brandl and
                  Matthias Wenzel and
                  Alessandro B. Melchiorre and
                  Elisabeth Lex and
                  Navid Rekabsaz and
                  Markus Schedl},
  editor       = {Himan Abdollahpouri and
                  Shaghayegh Sahebi and
                  Mehdi Elahi and
                  Masoud Mansoury and
                  Babak Loni and
                  Zahra Nazari and
                  Maria Dimakopoulou},
  title        = {Exploring Cross-group Discrepancies in Calibrated Popularity for Accuracy/Fairness
                  Trade-off Optimization},
  booktitle    = {Proceedings of the 2nd Workshop on Multi-Objective Recommender Systems
                  co-located with 16th {ACM} Conference on Recommender Systems (RecSys
                  2022), Seattle, WA, USA, 18th-23rd September 2022},
  series       = {{CEUR} Workshop Proceedings},
  volume       = {3268},
  publisher    = {CEUR-WS.org},
  year         = {2022},
  url          = {https://ceur-ws.org/Vol-3268/paper5.pdf},
  bibsource    = {dblp computer science bibliography, https://dblp.org}
}

@article{DBLP:journals/umuai/DeldjooJBDZ24,
  author       = {Yashar Deldjoo and
                  Dietmar Jannach and
                  Alejandro Bellog{\'{\i}}n and
                  Alessandro Difonzo and
                  Dario Zanzonelli},
  title        = {Fairness in recommender systems: research landscape and future directions},
  journal      = {User Model. User Adapt. Interact.},
  volume       = {34},
  number       = {1},
  pages        = {59--108},
  year         = {2024},
  url          = {https://doi.org/10.1007/s11257-023-09364-z},
  doi          = {10.1007/S11257-023-09364-Z},
  bibsource    = {dblp computer science bibliography, https://dblp.org}
}

@article{DBLP:journals/ftir/Ekstrand0B022,
  author       = {Michael D. Ekstrand and
                  Anubrata Das and
                  Robin Burke and
                  Fernando Diaz},
  title        = {Fairness in Information Access Systems},
  journal      = {Found. Trends Inf. Retr.},
  volume       = {16},
  number       = {1-2},
  pages        = {1--177},
  year         = {2022},
  url          = {https://doi.org/10.1561/1500000079},
  doi          = {10.1561/1500000079},
  bibsource    = {dblp computer science bibliography, https://dblp.org}
}

@inproceedings{DBLP:conf/um/SchedlRLGG22,
  author       = {Markus Schedl and
                  Navid Rekabsaz and
                  Elisabeth Lex and
                  Tessa Grosz and
                  Elisabeth Greif},
  title        = {Multiperspective and Multidisciplinary Treatment of Fairness in Recommender
                  Systems Research},
  booktitle    = {{UMAP} '22: 30th {ACM} Conference on User Modeling, Adaptation and
                  Personalization, Barcelona, Spain, July 4 - 7, 2022, Adjunct Proceedings},
  pages        = {90--94},
  publisher    = {{ACM}},
  year         = {2022},
  url          = {https://doi.org/10.1145/3511047.3536400},
  doi          = {10.1145/3511047.3536400},
  bibsource    = {dblp computer science bibliography, https://dblp.org}
}

@inproceedings{DBLP:conf/chiir/FerwerdaIBS23,
  author       = {Bruce Ferwerda and
                  Eveline Ingesson and
                  Michaela Berndl and
                  Markus Schedl},
  editor       = {Jacek Gwizdka and
                  Soo Young Rieh},
  title        = {I Don't Care How Popular You Are! Investigating Popularity Bias in
                  Music Recommendations from a User's Perspective},
  booktitle    = {Proceedings of the 2023 Conference on Human Information Interaction
                  and Retrieval, {CHIIR} 2023, Austin, TX, USA, March 19-23, 2023},
  pages        = {357--361},
  publisher    = {{ACM}},
  year         = {2023},
  url          = {https://doi.org/10.1145/3576840.3578287},
  doi          = {10.1145/3576840.3578287},
  bibsource    = {dblp computer science bibliography, https://dblp.org}
}

@article{BauerSchedlMainstream,
    doi = {10.1371/journal.pone.0217389},
    author = {Bauer, Christine AND Schedl, Markus},
    journal = {PLOS ONE},
    publisher = {Public Library of Science},
    title = {Global and country-specific mainstreaminess measures: Definitions, analysis, and usage for improving personalized music recommendation systems},
    year = {2019},
    volume = {14},
    pages = {1-36},
    number = {6},
}

@inproceedings{DBLP:conf/recsys/UngruhDVPH24,
  author       = {Robin Ungruh and
                  Karlijn Dinnissen and
                  Anja Volk and
                  Maria Soledad Pera and
                  Hanna Hauptmann},
  editor       = {Tommaso Di Noia and
                  Pasquale Lops and
                  Thorsten Joachims and
                  Katrien Verbert and
                  Pablo Castells and
                  Zhenhua Dong and
                  Ben London},
  title        = {Putting Popularity Bias Mitigation to the Test: {A} User-Centric Evaluation
                  in Music Recommenders},
  booktitle    = {Proceedings of the 18th {ACM} Conference on Recommender Systems, RecSys
                  2024, Bari, Italy, October 14-18, 2024},
  pages        = {169--178},
  publisher    = {{ACM}},
  year         = {2024},
  url          = {https://doi.org/10.1145/3640457.3688102},
  doi          = {10.1145/3640457.3688102},
  bibsource    = {dblp computer science bibliography, https://dblp.org}
}

@inproceedings{DBLP:conf/um/AlvesJSM24,
  author       = {Gabrielle Alves and
                  Dietmar Jannach and
                  Rodrigo Ferrari de Souza and
                  Marcelo Garcia Manzato},
  title        = {User Perception of Fairness-Calibrated Recommendations},
  booktitle    = {Proceedings of the 32nd {ACM} Conference on User Modeling, Adaptation
                  and Personalization, {UMAP} 2024, Cagliari, Italy, July 1-4, 2024},
  pages        = {78--88},
  publisher    = {{ACM}},
  year         = {2024},
  url          = {https://doi.org/10.1145/3627043.3659558},
  doi          = {10.1145/3627043.3659558},
  bibsource    = {dblp computer science bibliography, https://dblp.org}
}

@inproceedings{DBLP:conf/recsys/DokoupilBP25,
  author       = {Patrik Dokoupil and
                  Ludovico Boratto and
                  Ladislav Peska},
  editor       = {M{\'{a}}ria Bielikov{\'{a}} and
                  Pavel Kord{\'{\i}}k and
                  Markus Schedl and
                  Marco de Gemmis and
                  Sole Pera and
                  Rodrigo Alves and
                  Olivier Jeunen and
                  Vito Ostuni},
  title        = {How Do Users Perceive Recommender Systems' Objectives?},
  booktitle    = {Proceedings of the Nineteenth {ACM} Conference on Recommender Systems,
                  RecSys 2025, Prague, Czech Republic, September 22-26, 2025},
  pages        = {165--176},
  publisher    = {{ACM}},
  year         = {2025},
  url          = {https://doi.org/10.1145/3705328.3748066},
  doi          = {10.1145/3705328.3748066},
  bibsource    = {dblp computer science bibliography, https://dblp.org}
}

@article{DBLP:journals/tors/SilvaJ26,
  author       = {Diego Corr{\^{e}}a da Silva and
                  Dietmar Jannach},
  title        = {Calibrated Recommendations: Survey and Future Directions},
  journal      = {Trans. Recomm. Syst.},
  volume       = {4},
  number       = {3},
  pages        = {43:1--43:32},
  year         = {2026},
  url          = {https://doi.org/10.1145/3789266},
  doi          = {10.1145/3789266},
  bibsource    = {dblp computer science bibliography, https://dblp.org}
}

@inproceedings{Dokoupil2024,
author = {Dokoupil, Patrik and Boratto, Ludovico and Peska, Ladislav},
title = {User Perceptions of Diversity in Recommender Systems},
year = {2024},
isbn = {9798400704338},
publisher = {Association for Computing Machinery},
address = {New York, NY, USA},
url = {https://doi.org/10.1145/3627043.3659555},
doi = {10.1145/3627043.3659555},
booktitle = {Proceedings of the 32nd ACM Conference on User Modeling, Adaptation and Personalization},
pages = {212–222},
numpages = {11},
location = {Cagliari, Italy},
series = {UMAP '24}
}

@inproceedings{DBLP:conf/ht/dinnissen2025role,
  title={The Role of Fairness and Diversity in User Choices and Perceptions of Music Playlists},
  author={Dinnissen, Karlijn and Khan, Shah Noor and Hauptmann, Hanna and Herder, Eelco and Masthoff, Judith},
  booktitle={Proceedings of the 36th ACM Conference on Hypertext and Social Media},
  pages={48--57},
  year={2025}
}

@inproceedings{DBLP:conf/rs/liang2022exploring,
  title={Exploring the longitudinal effects of nudging on users’ music genre exploration behavior and listening preferences},
  author={Liang, Yu and Willemsen, Martijn C},
  booktitle={Proceedings of the 16th ACM conference on recommender systems},
  pages={3--13},
  year={2022}
}

@article{porcaro2024assessing,
  title={Assessing the impact of music recommendation diversity on listeners: A longitudinal study},
  author={Porcaro, Lorenzo and G{\'o}mez, Emilia and Castillo, Carlos},
  journal={ACM Transactions on Recommender Systems},
  volume={2},
  number={1},
  pages={1--47},
  year={2024},
  publisher={ACM New York, NY}
}

@inproceedings{DBLP:conf/recsys/LesotaBWWTS25,
  author       = {Oleg Lesota and
                  Adrian Bajko and
                  Max Walder and
                  Matthias Wenzel and
                  Antonela Tommasel and
                  Markus Schedl},
  editor       = {M{\'{a}}ria Bielikov{\'{a}} and
                  Pavel Kord{\'{\i}}k and
                  Markus Schedl and
                  Marco de Gemmis and
                  Sole Pera and
                  Rodrigo Alves and
                  Olivier Jeunen and
                  Vito Ostuni},
  title        = {Fine-tuning for Inference-efficient Calibrated Recommendations},
  booktitle    = {Proceedings of the Nineteenth {ACM} Conference on Recommender Systems,
                  RecSys 2025, Prague, Czech Republic, September 22-26, 2025},
  pages        = {1187--1192},
  publisher    = {{ACM}},
  year         = {2025},
  url          = {https://doi.org/10.1145/3705328.3759319},
  doi          = {10.1145/3705328.3759319},
  bibsource    = {dblp computer science bibliography, https://dblp.org}
}

\end{document}